\newif\ifincludesupplement
\includesupplementtrue

\newif\ifjaisclass
\IfFileExists{template/phep.cls}{\jaisclasstrue}{\jaisclassfalse}
\ifjaisclass
  \documentclass{template/phep}
\else
  \documentclass[10pt,twocolumn]{article}
  \usepackage[a4paper,margin=18mm]{geometry}
  \newcommand{\journal}[1]{}
  \newcommand{\vol}[1]{}
  \newcommand{\auno}[1]{\textsuperscript{#1}}
  \newcommand{\address}[1]{\date{\parbox{0.96\textwidth}{\small #1}}}
  \newcommand{\tablecaptionfont}{\small}
\fi

\journal{JAIS}
\vol{Preprint}
\usepackage{amsmath}
\usepackage{amssymb}
\usepackage{booktabs}
\usepackage{graphicx}
\usepackage[hidelinks]{hyperref}
\usepackage{tikz}
\usetikzlibrary{arrows.meta,positioning}
\usepackage{placeins}
\usepackage{stfloats}

\ifjaisclass\else\fi
\newcommand{\Ecrit}{E_{\mathrm{crit}}}
\newcommand{\Nsource}{N_{\mathrm{source}}}

\newcommand{\paperabstract}{%
Transmission muography usually treats each angular bin as a straight line of sight through the target. Multiple Coulomb scattering can break this mapping by changing the muon direction in rock. We introduce CABRIALES, a reduced model that represents this effect as migration between source and destination angular bins. The input muons are generated with a conditional normalizing flow trained on ARTI simulations, selected with a continuous-slowing-down range threshold, and deflected using angular distributions derived from Geant4. We apply the model to four observation points at Cerro Mach\'{i}n volcano in Colombia. Among the muons selected by the transmission threshold, 41.4--45.1\% move to a different $2.5^{\circ}$ bin within the field of view, while 20.7--24.3\% reach a different rock-intersecting angular bin associated with a longer rock path in the terrain model. For muons with energies within 5\% above the transmission threshold, the changed-bin fraction increases to 67.4--81.3\%. Destination maps further show that events arriving from shorter assigned paths can form a large share of the modeled population in some rock-intersecting bins. These results quantify angular redistribution before detector response; they do not represent net count changes or density bias.}

\newcommand{\paperkeywords}{%
\par\vspace{8pt}
\noindent\textit{Keywords:} muography, multiple Coulomb scattering, angular migration, volcanic imaging, Cerro Mach\'{i}n}

\begin{document}

\title{Simulation-Derived Angular Migration in Volcanic Muography}
\author{%
R.A.~Mart\'inez-Rivero\auno{1,*},
C.~Sarmiento-Cano\auno{1,2}, and
L.A.~N\'u\~nez\auno{1,3}}
\address{%
\begin{flushleft}
$^1$Escuela de F\'isica, Universidad Industrial de Santander, Bucaramanga, Colombia.\\
$^2$Departamento de Ciencias B\'asicas, Universidad Aut\'onoma de Bucaramanga, Bucaramanga, Colombia.\\
$^3$Departamento de F\'isica, Universidad de Los Andes, M\'erida, Venezuela.
\end{flushleft}}

\ifjaisclass
\begin{abstract}
\begin{flushleft}\paperabstract\paperkeywords\end{flushleft}
\end{abstract}
\maketitle
\else
\maketitle
\begin{abstract}\paperabstract\paperkeywords\end{abstract}
\fi

\renewcommand{\thefootnote}{\fnsymbol{footnote}}
\footnotetext[1]{Corresponding author: rafael2248058@correo.uis.edu.co}

\section{Introduction}

Transmission muography exploits the attenuation of atmospheric muons to estimate the opacity of geological structures, defined as the density integrated along the muon path~\cite{Bae2026,Tanaka2007,Lesparre2010}. A digital elevation model (DEM) provides the rock path length associated with each observation direction. In the conventional ballistic approximation, muons are assumed to travel along straight paths without changing direction. Consequently, each incoming angular bin contributes only to its corresponding outgoing bin.

In practice, multiple Coulomb scattering (MCS) deflects muons as they traverse rock, with larger angular deviations occurring at lower momenta~\cite{Highland1975,Lynch1991}. Scattering can therefore become particularly relevant for muons with energies close to the minimum required to cross a given rock path. As a result, a muon associated with a relatively short incoming path may emerge in an angular bin to which the DEM assigns a longer path. This angular reassignment does not mean that the muon has traversed the longer rock column, but it changes the angular bin to which the event is assigned.

Previous studies have used particle-transport simulations to investigate the effects of scattering on angular resolution, forward-scattering backgrounds, and upward-going particles in muography~\cite{Olah2019,Gomez2017,Jourde2013}. Some of these studies considered particles scattered into the detector from outside the target. Here we address a different effect: the redistribution of muons selected by the transmission model among finite angular bins. Fast simulation tools such as MUONITH have also introduced approximate transport methods for muography observation planning~\cite{MUONITH2026}. Our approach instead represents scattering-induced angular migration explicitly as a transfer between source and destination bins, before detector response is applied~\cite{MuTeResponse}.

To quantify this effect, we introduce CABRIALES, a computational model of angular migration in transmission muography. For each input muon, the model uses the initial energy, incoming direction, and DEM-assigned rock path length to select the source population with a continuous slowing down approximation (CSDA) transmission threshold. It then samples an outgoing direction from angular-deflection distributions derived from Geant4 simulations. This provides a reduced description of angular redistribution without tracking complete particle trajectories through the three-dimensional terrain.

In this work, we apply CABRIALES to four observation points at Cerro Mach\'{i}n volcano in Colombia. By retaining both the source and destination bins on a $2.5^\circ\times2.5^\circ$ grid, the analysis measures the fraction of CSDA-selected muons that change angular bin, migrate to rock-intersecting directions with longer DEM-assigned paths, or leave the angular field of view (FOV). We also examine how these quantities depend on the muon energy relative to the transmission threshold.

\section{Angular-migration model}

\subsection{From source bins to destination bins}

In the ballistic approximation, the source population consists of muons whose initial energy is high enough to cross the DEM-assigned rock path along their incoming direction. For source bin $i$, we denote this population by $N_i^{\mathrm{bal}}$, where ``bal'' refers to the ballistic assignment. Multiple Coulomb scattering redistributes these muons among destination bins $j$. We write this angular transfer as
\begin{equation}
N_j^{\mathrm{out}}=\sum_i\overline{M}_{ji}N_i^{\mathrm{bal}}.
\label{eq:matrix}
\end{equation}
Here, $N_j^{\mathrm{out}}$ is the population assigned to destination bin $j$ after scattering. Each element $\overline{M}_{ji}$ is the source-weighted probability that an event from bin $i$ is assigned to bin $j$. Without scattering, only diagonal elements contribute and the outgoing angular assignment is identical to the ballistic one.

The source population is selected with the CSDA transmission threshold
\begin{equation}
\tau_{\mathrm{CSDA}}(E_\mu,\Omega)=
\mathbf{1}\!\left[E_\mu\geq\Ecrit\bigl(L(\Omega)\bigr)\right].
\label{eq:csda-selection}
\end{equation}
The function is one when the event satisfies the threshold and zero otherwise. Here, $E_\mu$ is the initial kinetic energy, $\Omega$ is the incoming direction, and $L(\Omega)$ is the DEM-assigned rock path length. The quantity $\Ecrit(L)$ is the minimum kinetic energy given by the SiO$_2$ CSDA range relation used in this work (Supplementary Section~S1). This deterministic criterion defines the source sample; stochastic fluctuations in energy loss are not included in the selection.

Because $\overline{M}_{ji}$ averages over the selected muons in source bin $i$, it depends on their energy and angular distribution. For each nonempty source bin, the probabilities over all possible destinations sum to one.

\subsection{Angular deflections}

Once a muon satisfies the transmission criterion in Eq.~\eqref{eq:csda-selection}, CABRIALES must assign its outgoing direction. The required angular-deflection distributions are obtained from Geant4 simulations of muons crossing homogeneous SiO$_2$ layers~\cite{Agostinelli2003,Allison2016}. For each simulated energy and layer thickness, the muons that leave the material define one-dimensional distributions of projected angular deflections.

For each selected muon, CABRIALES samples one projected deflection along each of two transverse directions and uses the two values to update the direction on the unit sphere. The two components are sampled independently. This approximation simplifies the lookup procedure but does not preserve possible correlations between the transverse components. The detailed Geant4 configuration, library coverage, interpolation rules, and vector update are given in the supplementary methods.

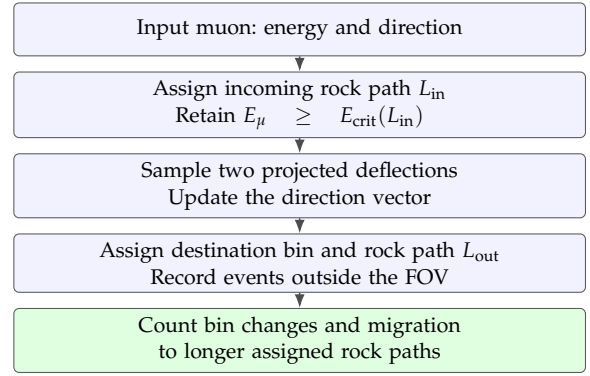
\begin{figure}[!htbp]
\centering
\begin{tikzpicture}[
 node distance=2.0mm,
 box/.style={draw=black!65,rounded corners=2pt,fill=blue!5,
 text width=0.84\columnwidth,minimum height=7mm,align=center,font=\small},
 arrow/.style={-{Latex[length=1.8mm]},semithick,draw=black!70}]
\node[box] (source) {Input muon: energy and direction};
\node[box,below=of source] (select) {Assign incoming rock path $L_{\mathrm{in}}$\\Retain $E_\mu\geq\Ecrit(L_{\mathrm{in}})$};
\node[box,below=of select] (scatter) {Sample two projected deflections\\Update the direction vector};
\node[box,below=of scatter] (dest) {Assign destination bin and rock path $L_{\mathrm{out}}$\\Record events outside the FOV};
\node[box,below=of dest,fill=green!12] (count) {Count bin changes and migration\\to longer assigned rock paths};
\draw[arrow] (source)--(select);
\draw[arrow] (select)--(scatter);
\draw[arrow] (scatter)--(dest);
\draw[arrow] (dest)--(count);
\end{tikzpicture}
\caption{Single-muon calculation in CABRIALES. Each input muon is selected by the CSDA threshold for its incoming DEM-assigned rock path. The model then samples two projected angular deflections, updates the direction, and records the source and destination bins and their assigned path lengths. The calculation does not track a three-dimensional trajectory through the terrain.}
\label{fig:workflow}
\end{figure}

The source selection and angular redistribution are therefore treated separately: the CSDA criterion decides which input muons enter the calculation, while the Geant4-derived distributions determine how their directions change. The reported fractions describe this reduced angular model rather than complete stochastic transport through Cerro Mach\'{i}n. Figure~\ref{fig:workflow} summarizes the calculation.

\subsection{Migration observables}

After sampling the outgoing direction, we compare the source and destination bins. The first observable is the fraction of selected events that move to another bin while remaining inside the FOV:
\begin{equation}
P_{\mathrm{chg}}=\frac{1}{\Nsource}\sum_k
\mathbf{1}\!\left[i_k\ne j_k,\ \Omega_{\mathrm{out},k}\in\mathrm{FOV}\right].
\label{eq:changed}
\end{equation}
Here, $i_k$ and $j_k$ are the source and destination bins of event $k$, and $\Nsource$ is the number of selected source events. Events that leave the FOV remain in the denominator and are recorded separately as overflow. Because the initial direction can lie close to a bin boundary, changing bin does not require a deflection as large as the bin width.

The FOV contains directions assigned to mapped rock as well as directions that do not intersect mapped rock. The second observable therefore restricts the destination to a rock direction and asks whether scattering moves the event to a different bin with a longer DEM-assigned path:
\begin{equation}
\begin{split}
P_{>}={}&\frac{1}{\Nsource}\sum_k\mathbf{1}\!\bigl[
 i_k\ne j_k,\ \Omega_{\mathrm{out},k}\in\mathrm{FOV},\\
&s_{\mathrm{out},k}=\mathrm{rock},\
 L_{\mathrm{out},k}>L_{\mathrm{in},k}\bigr].
\end{split}
\label{eq:strict}
\end{equation}
Here, $L_{\mathrm{in},k}\equiv L(\Omega_{\mathrm{in},k})$ and $L_{\mathrm{out},k}\equiv L(\Omega_{\mathrm{out},k})$ are the DEM-assigned path lengths for the incoming and outgoing directions. The class $s_{\mathrm{out},k}=\mathrm{rock}$ identifies a destination direction that intersects mapped rock. These lengths are straight-line DEM assignments, not the length of a scattered trajectory through the terrain. At fixed density, a larger assigned path also corresponds to a larger assigned opacity.

Both observables are normalized to the selected source population. When we restrict the analysis to an energy interval, the denominator is the selected source count in that interval.

\begin{figure*}[!tbp]
\centering
\includegraphics[width=0.88\textwidth]{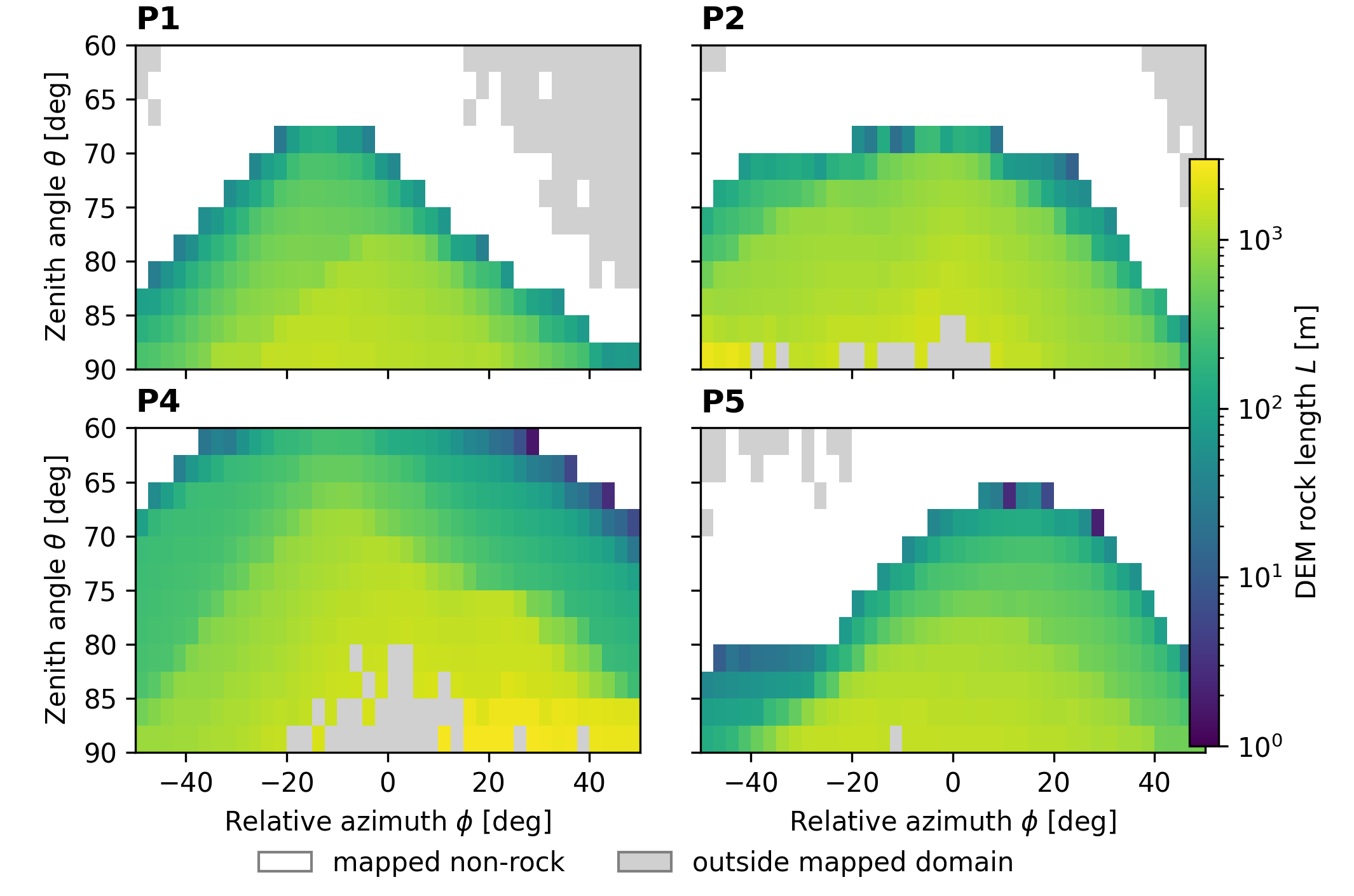}
\caption{DEM rock-length maps for P1, P2, P4, and P5 (Table~\ref{tab:geometry}), displayed on a $2.5^\circ$ grid with a common logarithmic scale. White cells are mapped non-rock directions. Gray cells lack a mapped DEM assignment; they do not denote FOV overflow.}
\label{fig:dem-length}
\end{figure*}

\section{Application to Cerro Mach\'{i}n}

We apply the same migration model to four observation points at Cerro Mach\'{i}n volcano in Colombia, denoted P1, P2, P4, and P5 throughout this work. Table~\ref{tab:geometry} gives their coordinates and selected event counts, while Fig.~\ref{fig:dem-length} shows the corresponding DEM-assigned rock paths. The corresponding observation geometries have different angular boundaries and rock-length patterns, allowing us to compare migration across distinct observation geometries.

\begin{table}[!htbp]
\centering\small
\refstepcounter{table}\label{tab:geometry}
\parbox{\columnwidth}{\tablecaptionfont\textbf{Table \thetable:} Observation locations for the four observation points. Coordinates are in decimal degrees. $\Nsource$ counts events after angular, rock-source, and CSDA selection.\par}
\vspace{4pt}
\setlength{\tabcolsep}{4pt}
\begin{tabular}{lrrr}
\toprule
Observation point & Latitude & Longitude & $\Nsource$\\
\midrule
P1 & 4.492298 & -75.381092 & 168,893\\
P2 & 4.494946 & -75.388110 & 149,677\\
P4 & 4.476500 & -75.386500 & 816,575\\
P5 & 4.488500 & -75.379500 & 230,570\\
\bottomrule
\end{tabular}
\end{table}

\subsection{Muon source}

The source generator is based on atmospheric-muon simulations produced with the ARTI framework~\cite{ARTI2022}. ARTI uses CORSIKA to simulate cosmic-ray showers under site-specific atmospheric and geomagnetic conditions and records the secondary particles that reach the observation level. We have previously used ARTI-based muon simulations in detector and imaging studies~\cite{Hydroreactor}.

For this work, a conditional normalizing flow (CNF) trained on ARTI simulations generates the input muons. The CNF samples muon energy and zenith angle jointly. CABRIALES then maps these particles to each observation point and applies the CSDA selection defined in Eq.~\eqref{eq:csda-selection}.

Across the four observation points, 1,365,715 events satisfy the source selection. We report fractions of this population rather than exposure-normalized fluxes. The same generated atmospheric source is evaluated for each observation point, so pooled quantities sum the selected event counts across observation points rather than treating them as independent atmospheric realizations. P4 contains the largest selected sample (59.79\% of the pooled events) and also spans a broad mapped rock domain in Fig.~\ref{fig:dem-length}.

\subsection{Angular grid and DEM assignments}

All central results use the angular window $60^\circ\leq\theta<90^\circ$ and $-50^\circ\leq\phi<50^\circ$, divided into $2.5^\circ\times2.5^\circ$ bins. The FOV and the mapped DEM domain are not identical. Within the angular window, some directions intersect mapped rock, some are mapped as non-rock, and some have no valid DEM assignment. CABRIALES keeps this geometric classification separate from the coarse angular bins. The detailed bin convention and coordinate definitions are given in the supplementary methods.

\section{Results}

\subsection{Bin migration and energy dependence}

Table~\ref{tab:main} gives the central result. Across the four observation points, 41.4--45.1\% of the selected events move to a different $2.5^\circ$ bin while remaining inside the FOV. A subset of 20.7--24.3\% reaches a different rock bin with a longer DEM-assigned path. The pooled fractions are 42.10\% and 23.17\%, respectively. Including events that leave the FOV gives a changed-or-overflow fraction of 47.6\%.

P4 differs from the other observation points mainly through its overflow fraction. Its mapped rock domain reaches several boundaries of the FOV (Fig.~\ref{fig:dem-length}), which is consistent with a larger fraction of scattered events leaving the angular window.

\begin{table*}[!tbp]
\centering\small
\refstepcounter{table}\label{tab:main}
\parbox{\textwidth}{\tablecaptionfont\textbf{Table \thetable:} Migration fractions (\%). ``Different bin'' is $P_{\mathrm{chg}}$ [Eq.~\eqref{eq:changed}]. ``Longer assigned path'' is $P_{>}$ [Eq.~\eqref{eq:strict}]. These columns and overflow use $\Nsource$ from Table~\ref{tab:geometry}. ``Near threshold'' is the changed-bin fraction for $1\leq E_\mu/\Ecrit<1.05$. Its source counts are 8,840, 8,373, 39,293, and 11,493 for P1, P2, P4, and P5, respectively (67,999 pooled). Pooled values use the corresponding counts, not equal weights for the observation points.\par}
\vspace{4pt}
\setlength{\tabcolsep}{9pt}
\begin{tabular}{lrrrr}
\toprule
Observation point & Different bin & Longer assigned path & Overflow & Near threshold\\
\midrule
P1 & 41.38 & 20.74 & 0.62 & 79.49\\
P2 & 45.09 & 22.33 & 0.65 & 81.29\\
P4 & 41.87 & 24.27 & 8.67 & 67.44\\
P5 & 41.53 & 21.59 & 0.95 & 79.13\\
\midrule
Pooled & 42.10 & 23.17 & 5.49 & 72.69\\
\bottomrule
\end{tabular}
\end{table*}

\begin{figure}[!htbp]
\centering
\includegraphics[width=\columnwidth]{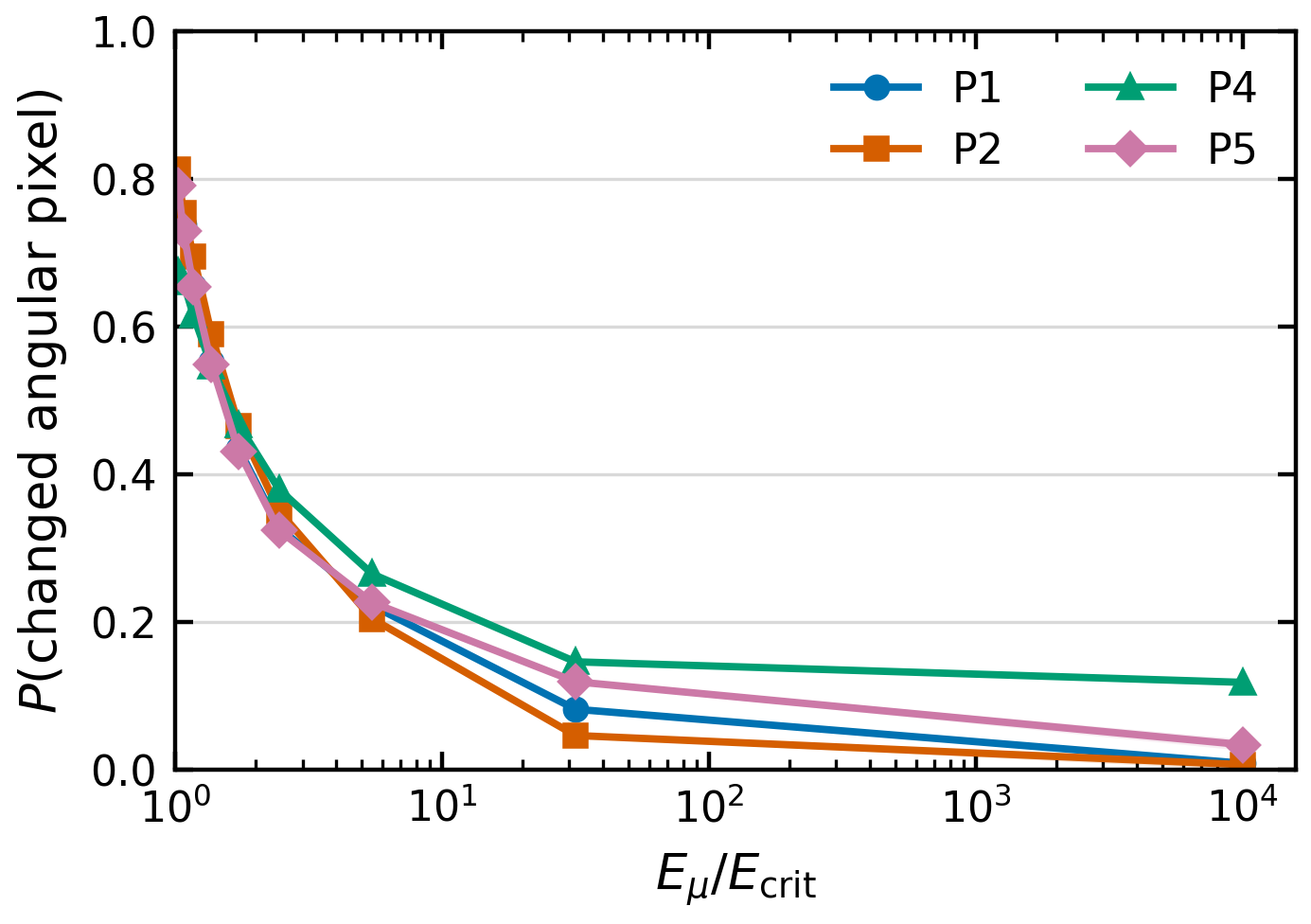}
\caption{Changed-bin fraction $P_{\mathrm{chg}}$ within each $E_\mu/\Ecrit$ interval (labeled as ``changed angular pixel'' on the vertical axis). The final marker represents $[100,10^6)$ and is placed at $10^4$ for display. The error bars are approximate counting intervals for a fixed source sample, not total uncertainties. Lines guide the eye.}
\label{fig:energy}
\end{figure}

The migration probability increases near the CSDA threshold (Fig.~\ref{fig:energy}). For $1\leq E_\mu/\Ecrit<1.05$, 67.4--81.3\% of the selected events change bin. P4 shows the lowest near-threshold in-FOV changed-bin fraction among the four observation points (67.44\%, Table~\ref{tab:main}). Events below $1.25\Ecrit$ represent 20.8\% of the source but 33.2\% of all in-FOV bin changes. Near-threshold events therefore contribute disproportionately to the angular redistribution predicted by the model.

This trend is qualitatively consistent with the larger multiple-scattering deflections expected at lower muon momentum~\cite{Highland1975,Lynch1991}. Its magnitude is not determined by $E_\mu/\Ecrit$ alone: absolute energy, assigned path length, and the initial position within an angular bin also vary. Each point in Fig.~\ref{fig:energy} therefore represents an average over the events in that interval.

\subsection{Assigned path length and destination populations}

The DEM path length varies continuously with direction, so a deflected muon can reach a direction with a longer assigned path without leaving its original $2.5^\circ$ bin. In the pooled sample, 41.2\% of events reach a direction with a larger DEM-assigned path. Of these, 18.0\% remain inside the same coarse bin, while 23.17\% also migrate to a different rock bin. The inter-bin migration is not limited to very small path changes: increases larger than 10, 50, and $100~\mathrm{m}$ occur for 20.4\%, 12.8\%, and 7.9\% of the selected sample, respectively.

\begin{figure*}[!tbp]
\centering
\begin{tikzpicture}
\node[anchor=south west,inner sep=0] (image) at (0,0)
{\includegraphics[width=0.94\textwidth]{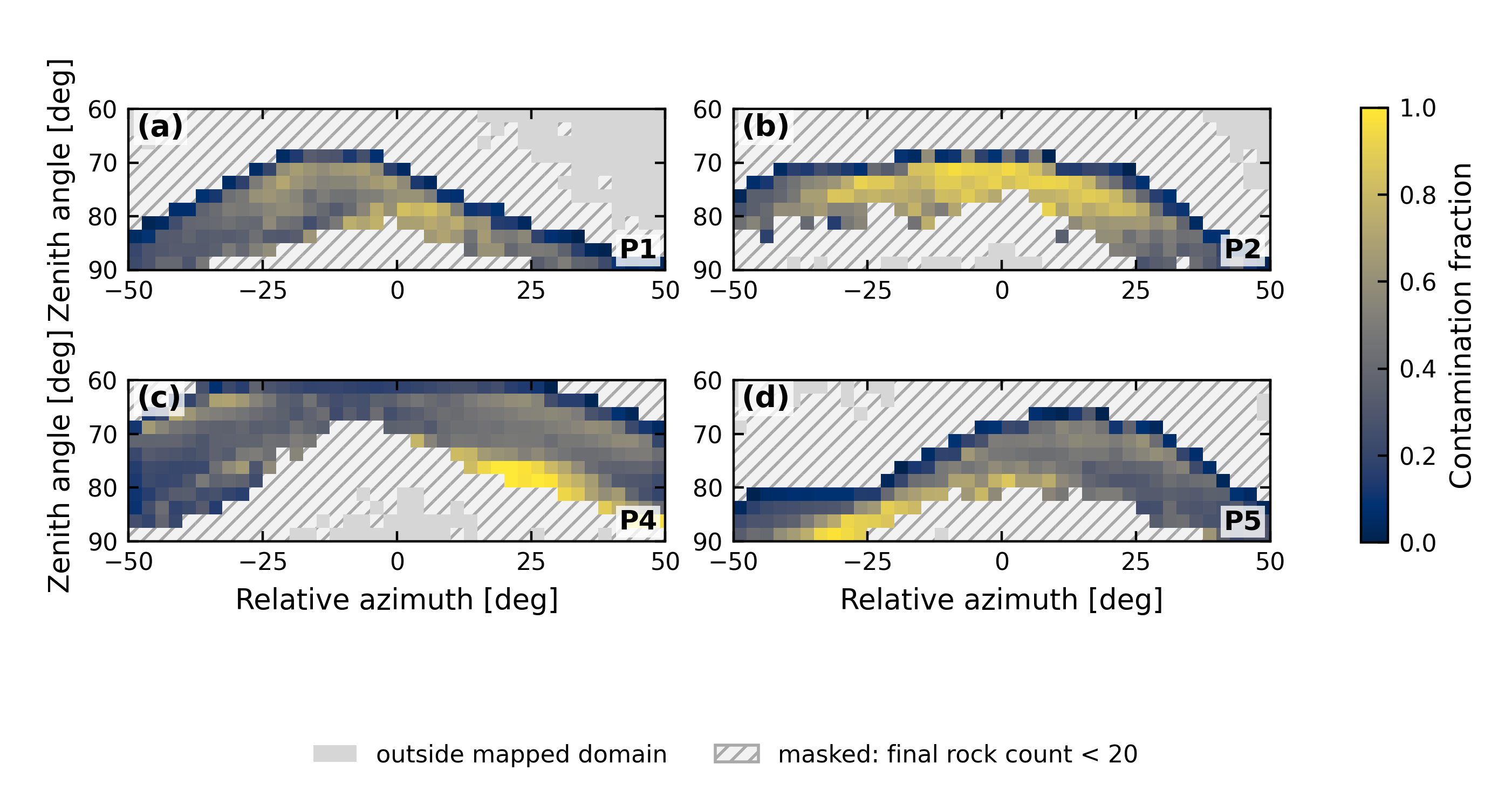}};
\begin{scope}[x={(image.south east)},y={(image.north west)}]
\fill[white] (0.956,0.21) rectangle (0.998,0.88);
\node[rotate=90,font=\footnotesize] at (0.974,0.585)
{Lower-opacity contribution fraction};
\end{scope}
\end{tikzpicture}
\caption{Fraction of events in each rock destination bin that arrived from a different bin with a shorter assigned path [Eq.~\eqref{eq:lower-contribution}]. At fixed density this is also a contribution from lower assigned opacity. Gray cells lack a mapped DEM assignment; hatching masks bins with fewer than 20 final rock events. This is neither a measured detector background nor a net count enhancement.}
\label{fig:contamination}
\end{figure*}

The source-normalized quantity $P_{>}$ describes how often events move to a different rock bin with a longer assigned path. A complementary question is how much of the modeled population arriving in a given destination bin originated from shorter-path source bins. For each rock destination bin $j$, we define
\begin{equation}
f_{\mathrm{lower}\rightarrow j}=
\frac{N_{\mathrm{lower}\rightarrow j}}
{N^{\mathrm{out}}_{j,\mathrm{rock}}}.
\label{eq:lower-contribution}
\end{equation}
Here, $N_{\mathrm{lower}\rightarrow j}$ is the number of modeled events that arrive in bin $j$ from a different source bin with $L_{\mathrm{in}}<L_{\mathrm{out}}$, and $N^{\mathrm{out}}_{j,\mathrm{rock}}$ is the total modeled population assigned to rock bin $j$. Figure~\ref{fig:contamination} maps this destination-side fraction. Comparing Figs.~\ref{fig:dem-length} and~\ref{fig:contamination} shows a clear spatial structure: regions close to the upper boundary of the mapped rock domain generally receive a small contribution from source bins with shorter assigned paths, whereas several regions associated with larger DEM path lengths show substantially larger values of $f_{\mathrm{lower}\rightarrow j}$. In those regions, a large fraction of the modeled destination population originates from other angular bins with shorter assigned rock paths.

\section{Numerical checks and physical scope}

\subsection{Sensitivity to the transverse sampling basis}
\label{sec:numerical-checks}

CABRIALES samples the two transverse deflection components independently. This approximation can make the predicted migration depend on the orientation chosen for the local transverse axes. We test one such sensitivity by rotating the transverse basis by $45^\circ$ while keeping the same source events and sampled component values. The pooled changed-bin and overflow fractions change by $-0.15$ and $+0.23$ percentage points, respectively; the largest change among observation points is $0.72$ percentage points. These values describe one paired test for a fixed realization and one rotation. They do not establish rotational invariance or validate the full joint angular distribution against Geant4. Independent event recounts and boundary checks are documented in Supplementary Section~S3.

\subsection{Physical scope}

The reported fractions should be interpreted as predictions of the reduced angular-migration model. Two approximations are particularly important near the transmission threshold. First, the source is selected with a deterministic CSDA criterion, whereas the Geant4 deflection distributions are conditioned on simulated transmission through the reference material. A stochastic survival calculation could therefore change the mixture of events that contributes near threshold. Second, the two transverse deflection components are sampled independently, so the model does not preserve their full joint distribution.

The calculation also stops before the detector stage. It does not propagate the lateral displacement of the muon through the three-dimensional terrain, test whether the scattered trajectory intersects a finite detector, or apply detector acceptance and resolution~\cite{MuTeResponse}. The results therefore quantify angular redistribution before instrumental selection. Numerical interpolation, library coverage, extrapolation policies, and the available uncertainty diagnostics are documented in the supplementary methods.

\section{Conclusions}

We introduced CABRIALES to represent rock-induced angular scattering as migration between source and destination bins in transmission muography. Applied to four observation points at Cerro Mach\'{i}n, the model predicts that 41.4--45.1\% of CSDA-selected events move to another $2.5^\circ$ bin inside the FOV. A fraction of 20.7--24.3\% also reaches a different rock bin with a longer DEM-assigned path. The migration is strongest for events selected close to the CSDA threshold.

These results show that a purely ballistic bin assignment can miss substantial angular redistribution within the reduced model. The destination maps further show that contributions from shorter assigned paths are spatially nonuniform and can form a large share of the modeled population in some rock bins. The $45^\circ$ basis-rotation test produces small shifts for the tested realization, while not establishing general rotational invariance. Determining how this redistribution changes detected counts or reconstructed density requires stochastic survival, full geometry, and detector-response modeling beyond the scope of this proceeding.

\section*{Data and code availability}

The CABRIALES code is available at \url{https://github.com/rmartinezra/CABRIALES}. The base repository state is identified by commit \texttt{a571fe2}; the full identifier and the provenance of the event products used here are given in the supplementary methods. Full regeneration also requires the CNF source configuration and the original DEM-to-angle metadata. A persistent public archive has not yet been assigned.


\ifincludesupplement
\clearpage
\onecolumn
\begingroup
\pagestyle{plain}
\thispagestyle{plain}
\setlength{\parindent}{0pt}
\setlength{\parskip}{4pt}
\fontsize{10}{12}\selectfont

\pagenumbering{arabic}
\renewcommand{\thepage}{S\arabic{page}}
\setcounter{section}{0}
\setcounter{subsection}{0}
\setcounter{subsubsection}{0}
\setcounter{equation}{0}
\setcounter{table}{0}
\setcounter{figure}{0}
\setcounter{footnote}{0}
\renewcommand{\thesection}{S\arabic{section}}
\renewcommand{\thesubsection}{\thesection.\arabic{subsection}}
\renewcommand{\thesubsubsection}{\thesubsection.\arabic{subsubsection}}
\renewcommand{\theequation}{S\arabic{equation}}
\renewcommand{\thetable}{S\arabic{table}}
\renewcommand{\thefigure}{S\arabic{figure}}
\renewcommand{\thefootnote}{\arabic{footnote}}

\makeatletter
\@removefromreset{equation}{section}
\@removefromreset{table}{section}
\@removefromreset{figure}{section}
\makeatother

\providecommand{\theHsection}{}
\providecommand{\theHsubsection}{}
\providecommand{\theHsubsubsection}{}
\providecommand{\theHequation}{}
\providecommand{\theHtable}{}
\providecommand{\theHfigure}{}
\providecommand{\theHfootnote}{}
\renewcommand{\theHsection}{supp.\arabic{section}}
\renewcommand{\theHsubsection}{\theHsection.\arabic{subsection}}
\renewcommand{\theHsubsubsection}{\theHsubsection.\arabic{subsubsection}}
\renewcommand{\theHequation}{supp.\arabic{equation}}
\renewcommand{\theHtable}{supp.\arabic{table}}
\renewcommand{\theHfigure}{supp.\arabic{figure}}
\renewcommand{\theHfootnote}{supp.\arabic{footnote}}

\pdfbookmark[0]{Supplementary methods and numerical checks}{cabriales.supplement}
\begin{center}
{\Large\bfseries Supplementary methods and numerical checks\par}
\vspace{5pt}
{\large Simulation-Derived Angular Migration in Volcanic Muography\par}
\vspace{5pt}
R.A.~Mart\'inez-Rivero, C.~Sarmiento-Cano, and L.A.~N\'u\~nez
\end{center}
\vspace{4pt}

\section{The population-averaged migration matrix}

The main text starts with the transfer relation. Here we retain its integral definition, so that the source weighting and normalization remain explicit.

For source bin $i$, the ballistic spectrum is
\begin{equation}
\frac{dN_i^{\mathrm{bal}}}{dE_\mu}=
T_{\mathrm{live}}\int_{\Omega_i\cap\mathcal D_{\mathrm{rock}}}
 d\Omega\,\mathcal F(E_\mu,\Omega)\tau_{\mathrm{CSDA}}(E_\mu,\Omega),
\label{supp:eq:ballistic}
\end{equation}
where $\mathcal D_{\mathrm{rock}}$ contains the mapped rock source directions and
\begin{equation}
\tau_{\mathrm{CSDA}}(E_\mu,\Omega)=
\mathbf 1\!\left[E_\mu\geq\Ecrit\bigl(L(\Omega)\bigr)\right].
\end{equation}
The source-rate density $\mathcal F$ includes integration over the injection surface. Its units are energy$^{-1}$ time$^{-1}$ sr$^{-1}$. Thus, Eq.~\eqref{supp:eq:ballistic} has units of energy$^{-1}$. This normalization does not include detector acceptance or efficiency. Each direction keeps its own DEM length; a coarse bin is not assigned one common length.

The exposure $T_{\mathrm{live}}$ is part of the general rate notation. It does not assign a measured or verified simulated exposure to the archived sample. Equivalently, define the source number density $\mathcal G=T_{\mathrm{live}}\mathcal F$ and use $\mathcal G$ throughout. A common multiplicative normalization cancels from the fractions; a change in source shape does not generally cancel.

For each selected input state, the implemented sampler gives
\begin{equation}
Q_j(E_\mu,\Omega)=
P\!\left(\Omega_{\mathrm{out}}\in\Omega_j\mid
E_\mu,\Omega,L(\Omega);\mathcal K\right),
\end{equation}
where $\mathcal K$ denotes the library and its evaluation and sampling rules. The predicted population of destination bin $j$ is
\begin{equation}
N_j^{\mathrm{out}}=T_{\mathrm{live}}\sum_i
\int_{\Omega_i\cap\mathcal D_{\mathrm{rock}}}d\Omega\int dE_\mu\,
\mathcal F(E_\mu,\Omega)\tau_{\mathrm{CSDA}}(E_\mu,\Omega)Q_j(E_\mu,\Omega).
\label{supp:eq:out}
\end{equation}
For $N_i^{\mathrm{bal}}=\int dE_\mu\,dN_i^{\mathrm{bal}}/dE_\mu>0$, define
\begin{equation}
\overline M_{ji}=\frac{T_{\mathrm{live}}}{N_i^{\mathrm{bal}}}
\int_{\Omega_i\cap\mathcal D_{\mathrm{rock}}}d\Omega\int dE_\mu\,
\mathcal F(E_\mu,\Omega)\tau_{\mathrm{CSDA}}(E_\mu,\Omega)Q_j(E_\mu,\Omega).
\end{equation}
Equation~\eqref{supp:eq:out} then gives $N_j^{\mathrm{out}}=\sum_i\overline M_{ji}N_i^{\mathrm{bal}}$.
Columns for empty source bins need not be defined. For nonempty bins, including all destinations gives $\sum_j\overline M_{ji}=1$. A matrix restricted to the FOV need not sum to one. This is angular probability conservation for the selected population, not a test of stochastic survival.

\section{Numerical directions and deflection update}

The archived angles use degrees. The numerical unit vector is
\begin{equation}
\mathbf n=(\sin\theta\cos\phi,\sin\theta\sin\phi,\cos\theta).
\end{equation}
Thus, $\theta$ is measured from $+z$, and $\phi$ increases from $+x$ toward $+y$. For trigonometric evaluation, angles are converted to radians. With the local basis $\mathbf e_\theta,\mathbf e_\phi$ and sampled components $a,b$ in radians, the implemented update is
\begin{equation}
\mathbf n'=\cos r\,\mathbf n+
\frac{\sin r}{r}(a\mathbf e_\theta+b\mathbf e_\phi),
\qquad r=\sqrt{a^2+b^2}.
\label{supp:eq:update}
\end{equation}
The continuous value of $\sin r/r$ at zero is one. Archived kicks are in mrad and are divided by $10^3$ before using Eq.~\eqref{supp:eq:update}. The formula preserves the norm in exact arithmetic. It defines the implemented map; it does not prove that the joint distribution generated from independent projected components is physically correct, especially at large deflections.

The analysis FOV is $[60^\circ,90^\circ)\times[-50^\circ,50^\circ)$, with bin origin $(60^\circ,-50^\circ)$ and widths $(2.5^\circ,2.5^\circ)$. An internal edge belongs to the bin above it. Native source indices are DEM-cache indices, not the coarser analysis indices. The reduction recomputes both source and destination bins from their directions. It retains missing lengths as NaN and uses stored classes for the original destinations. A missing length does not imply sky.

The available source reader uses a mathematical azimuth and subtracts $\phi_0=90^\circ-A_{\mathrm{summit}}$. The DEM code uses north-clockwise azimuth with a positive relative-angle convention. Establishing the exact sign correspondence between these two relative-azimuth definitions requires the upstream DEM transformation inputs. The present analysis therefore retains the archived angular, length, and surface assignments without reinterpretation.

\section{Independent reduction of the archived events}

An independent verification pass over all 1,365,715 rows of the four archived event files, without removing events, retraining the CNF, or running new particle transport, reproduced all 1,944 summary quantities in the packaged numerical tables exactly.

\begin{table}[!htbp]
\centering
\refstepcounter{table}\par\noindent\parbox{\linewidth}{\small\textbf{Table \thetable:} Selected source populations. The near-threshold interval is $1\leq E_\mu/\Ecrit<1.05$.}\par\vspace{5pt}
\begin{tabular}{lrrr}
\toprule
Observation point & $\Nsource$ & Source share (\%) & Near-threshold count\\
\midrule
P1 & 168,893 & 12.3666 & 8,840\\
P2 & 149,677 & 10.9596 & 8,373\\
P4 & 816,575 & 59.7910 & 39,293\\
P5 & 230,570 & 16.8827 & 11,493\\
\midrule
Pooled & 1,365,715 & 100 & 67,999\\
\bottomrule
\end{tabular}
\end{table}

The pooled numerators are 575,028 in-FOV bin changes, 316,438 changes to a different rock bin with a longer assigned path, and 75,034 overflow events. Bin change and overflow are disjoint, so their union has 650,062 events (47.5987\%). The stored destination classes partition the sample as
\[
1,008,502\ \mathrm{rock}+282,179\ \mathrm{sky}+75,034\ \mathrm{overflow}
+0\ \mathrm{invalid}=1,365,715.
\]
The inequality $P_{>}\leq P_{\mathrm{chg}}$ holds in every checked group. IDs are unique within each event file. They are not original air-shower IDs and are not required to be unique across observation points.

In the near-threshold interval, 49,428 of 67,999 events change bin (72.6893\%) and 8,461 leave the FOV (12.4428\%). Events with $E_\mu/\Ecrit<1.25$ account for 284,736 of 1,365,715 inputs (20.8489\%) and 190,883 of 575,028 bin changes (33.1954\%). The pooled fractions always use summed numerators and denominators, not an unweighted average over observation points.

There are 373,985 rock destinations with changed native length inside the same coarse bin. Of these, 245,659 have a longer assigned path. The numbers with a different bin and $\Delta L>0,10,50,100~\mathrm m$ are 316,438, 278,751, 175,300, and 107,881. These counts separate within-bin length changes from inter-bin migration.

All retained events satisfy the documented CSDA selection. The reconstructed energy ratio, energy margin, and path-length increment agree with their stored values. Reinverting the range table using the serialized length produces threshold differences up to $2.01\times10^{-4}~\mathrm{GeV}$. They lie inside bounds derived from the documented float32 storage of length and total threshold energy, followed by subtraction of the muon rest energy. No fitted tolerance was introduced. This is a serialization check, not a survival validation.

\subsection{Angular reconstruction and zero-deflection test}

Equation~\eqref{supp:eq:update} reconstructs every stored outgoing direction and its native and $2.5^\circ$ bin/FOV classifications in the tested numerical environment. This establishes consistency with the archived kicks, not the identity of the historical generator executable.

With zero kicks, angular round-off is of order $10^{-14}$ degrees. Four P4 events lie at affected boundaries. One at $\phi=21^\circ$ changes its native bin; one at $\phi=20^\circ$ also changes its $2.5^\circ$ bin. Two at $\theta=60^\circ$ fall numerically below the lower FOV edge. Hence the $2.5^\circ$ null test has one bin change and two overflow events, rather than exact invariance of every classification. None belongs to the near-threshold interval. The reduction keeps the original IDs and coordinates without a boundary correction. These control counts are not subtracted from the original scattering results.

\subsection{One paired rotation of the transverse basis}

The test rotates only the transverse basis by $45^\circ$, leaving the same input directions, sampled component values, angular grid, and FOV. It does not generate a new source or new kicks. Table~\ref{supp:tab:rotation} gives the change from the original realization.

\begin{table}[!htbp]
\centering
\refstepcounter{table}\label{supp:tab:rotation}\par\noindent\parbox{\linewidth}{\small\textbf{Table \thetable:} Change in migration fractions after the $45^\circ$ basis rotation, in percentage points. ``Near'' denotes $1\leq E_\mu/\Ecrit<1.05$.}\par\vspace{5pt}
\begin{tabular}{lrrrr}
\toprule
Observation point & $\Delta P_{\mathrm{chg}}$ & $\Delta P_{\mathrm{overflow}}$ & $\Delta P_{\mathrm{chg}}^{\mathrm{near}}$ & $\Delta P_{\mathrm{overflow}}^{\mathrm{near}}$\\
\midrule
P1 & +0.272954 & -0.056249 & -0.226244 & +0.079186\\
P2 & +0.134289 & -0.060129 & +0.179147 & -0.191090\\
P4 & -0.403637 & +0.410495 & -0.531901 & +0.717685\\
P5 & +0.270634 & -0.029926 & +0.069608 & -0.008701\\
\midrule
Pooled & -0.147176 & +0.226841 & -0.302946 & +0.400006\\
\bottomrule
\end{tabular}
\end{table}

These are differences for a fixed finite realization. They have no attached confidence intervals and do not assume atmospheric independence. They neither prove dependence of the underlying law nor establish equivalence of bases. No new DEM lookup was performed. Consequently, the test does not evaluate a new $P_{>}$ or assign old lengths to rotated directions. It does not replace a comparison with joint two-dimensional Geant4 exit directions, which are not available in the packaged angular-library products used here.

\section{Library evaluation and statistical scope}

The packaged library has separate core and full-tail branches. The full-tail domain uses a hull in $\log_{10}L$ and $\log_{10}(E_\mu/T_{50})$; outside this domain, the sampler uses the core. Here $T_{50}$ is a library calibration associated with 50\% slab transmission. It is not the CSDA threshold $\Ecrit$ used to select the application source. The main-text energy ratio always uses the latter.

The documented core interpolation uses normalized one-dimensional histograms, interpolation in log-length and log-energy coordinates, and a nearest-configuration rule outside support. The tail-aware branch interpolates quantiles and combines the central distribution with tabulated tails. The available code uses a transition between 250 and 300 mrad. The core is embedded with zero numerical weight outside its tabulated support and normalized on the retained support. That is an implementation policy; the omitted physical tail is not measured to be zero.

In the later packaged application, the high-energy core policy rescales with the ratio of $1/(\beta p)$ values. It is used for 5.4\% of events. A nearest-configuration rule, labeled \texttt{nearest\_length\_fallback} outside supported lengths, is used for 2.7\%. The label does not by itself imply a neighbor search in length alone. The packaged interpolation and policy selector are not identical to the base-commit versions. The stored policy comparisons are internal comparisons, not independent Geant4 references. Their effects are not combined into a physical uncertainty.

Existing leave-one-out summaries give median absolute relative RMS errors of 2.7\% and 2.8\% for core and full-tail. Their 90th/95th percentiles are 25.3\%/37.3\% and 20.7\%/23.1\%, respectively. The complete upstream cross-validation constructors and held-out predictors are not available in the packaged materials used for this analysis. These values are therefore retained as diagnostics of the precalculated library rather than as an end-to-end uncertainty on the migration fractions.

The error bars in Fig.~\ref{fig:energy} are approximate Wilson counting intervals for the fixed source sample. Their nominal confidence level is not documented in the available analysis materials, so no confidence level is assigned here. The plotted intervals do not include uncertainty in the source generator, its training sample, the joint angular distribution, or the geographic transformation. Training on ARTI does not by itself quantify CNF approximation error. Joint source generation does not imply independent energy and zenith, and must not be confused with independent transverse scattering components.

\section{Provenance and remaining inputs}

The reference code state is commit
\begin{center}
\small\texttt{a571fe22c8d40a08362d8360397ee4952b12ff44}.
\end{center}
The event package \nolinkurl{cabriales_prism_handoff_sio2} and the reduction package \nolinkurl{cierre_minimo_20261009_eventos_archivados} are separate analysis snapshots. The closure manifest \nolinkurl{MANIFIESTO_CIERRE.json} identifies the reference code, packaged inputs, and reduction scripts by SHA-256. These identifiers specify the materials used for the present analysis.

The packaged SiO$_2$ CSDA range table records Geant4 11.2.2. Available angular-scattering logs for the main/tail subsets record Geant4 10.7.p04 and SiO$_2$ at $2.32~\mathrm{g\,cm^{-3}}$. Complete run metadata are not available for every upstream library subset, so a single Geant4 version is not assigned to the complete packaged angular library.

The application source is a joint energy--zenith CNF trained on ARTI muons; the available generator assigns azimuth uniformly. Reproducing the source generation from scratch requires the original CNF configuration or checkpoint together with its generation metadata. The present results use the archived generated events and report event fractions rather than an exposure-normalized flux.

The reduction uses the archived angular and DEM-length assignments and does not regenerate the DEM. Full geometric reconstruction requires the original angular-length grids, DEM provenance, observation orientations, and angle-transformation metadata. The operational numerical angle convention is documented in Section~S2. A persistent public archive identifier has not yet been assigned.

\clearpage
\endgroup
\fi
\end{document}